\documentclass[12pt]{article}
\usepackage[utf8]{inputenc}
\usepackage[a4paper, margin = 1in]{geometry}
\usepackage[backend=biber,sorting = none]{biblatex}
\usepackage[T1]{fontenc}
\usepackage{url}
\usepackage{graphicx}
\usepackage{geometry}
\usepackage{adjustbox}
\usepackage{caption}
\usepackage{subcaption}
\usepackage{float}
\usepackage{siunitx}
\usepackage[table]{xcolor}
\usepackage{multirow}
\usepackage{comment}
\usepackage{setspace}
\title{Investigating the role of visual experience with face-masks in face recognition during COVID-19}
\author{Srijita Karmakar, Koel Das}
\date{April 2022}

\begin{document}

\maketitle
\section*{Abstract}

The introduction of face masks during COVID-19 presents a potential challenge for human face perception and recognition. Face masks possibly hinder the holistic processing of faces leading to difficulty in facial recognition. Our present study aims to investigate this issue by probing the neuropsychological mechanisms of face recognition, while also exploring a possible learning effect observed in regularly seen (personally familiar) masked faces. Our stimuli consisted of personally familiar, famous, and unfamiliar Indian faces in masked and unmasked conditions. Subjects participated in a 2-back task wherein trials were balanced within and across blocks to represent all conditions identically, while behavioral and EEG responses were recorded.  Statistical analyses revealed significant main effects of familiarity and mask-conditions on performance accuracy and reaction-time (RT). The highest performance accuracy was observed for familiar and unmasked faces and the least for unfamiliar and masked ones, whereas RTs followed expected reverse trends. Notably, the difference in performance accuracy between unmasked and masked famous faces was more prominent than that for personally familiar faces. 
These observations are suggestive of a beneficial effect of visual experience found for frequently seen masked faces. Masked unfamiliar faces contributed to the highest proportion of false-positive errors, indicating the inherent difficulty in processing novel masked faces compared to known ones. EEG analysis revealed effect of face-masks on N250 for both famous and personally familiar faces and N170 for only personally familiar faces. Our study suggests that personal familiarity aids perceptual learning of masked faces, and this learning may not generalize across familiarity levels.

\section*{Introduction}

If the eyes are the windows to one's soul, the face is possibly a door open wide. Humans are extremely efficient in recognizing face and their face recognition ability is not hindered by low level visual changes including size, viewpoint, illumination and context(see \cite{johnston2009familiar,ramon2018familiarity} for review). Face recognition skill plays a crucial role in the context of animal interaction and communication \cite{perrett}, and the processing of familiar faces is remarkably superior compared to  unknown faces\cite{bruce1994stability,bruce2001matching,burton2015arguments,ramon2016real,gobbini2007neural}.

However, the advent of COVID-19 since beginning of 2020 brought about a major visuo-social shift in our daily life with respect to social interaction by introducing the face mask which was deemed mandatory by almost all countries. Since then, we have faced rapid and rampant adoption and integration of face-masks in our daily life. The use of face masks ushered an extensive alteration in the social interaction between humans. Given the importance of faces in human interaction and communication, it is natural to imagine how face-masks, which cover close to 50\% of a face, might affect face perception, recognition, and, on a larger-scale, human communication \cite{molnar}. Face-masks may have a potential negative impact on the usual face-recognition machinery that is at work in the human brain \cite{natoli}. Many studies have, in the time since the introduction of face-masks due to the pandemic, observed experimental findings and evidences to this effect\cite{carragher,noyes2021effect, gori,carbon,freud, baron}.
A behavioural study \cite{freud} has shown that face-masks hinder the subjects' performance in a cognitive-task (specifically, an adapted version of the Cambridge Face Memory Test \cite{cfmt}) which tested subject's memory of unmasked and masked faces. Face-masks not only limit abilities of face-recognition, but also severely impede emotion-recognition. Several studies have found a detrimental effect of face-masks on emotion and expression perception \cite{gori}, \cite{carbon}. Specifically, one study \cite{carbon} observed distinct confusion patterns in emotion-reading for masked faces (with many emotions being interpreted as neutral), while another study \cite{gori} observed that face-masks show especially pronounced impairment of emotion inference in children still in their developmental stages. Yet another study \cite{baron} has reported that face-masks affect social cognition in aged adults and people afflicted with dementia. Another recent  study found neural correlates of attentional processing in the presence of face-masks \cite{zochowska}. The face selective N170 component was found to be similar for both unmasked and masked faces, whereas the P100, P300 and LPP (attention-related components) were found to be higher for masked faces as opposed to unmasked faces.

Introduction of face mask not only impedes face perception and emotion recognition but arguably also impacts face familiarity. The first study \cite{carragher} exploring the effect of familiarity in masked-face perception found that face-masks cause a reduction in performance in human face-matching tasks. However, this detrimental effect was found to be similar for both familiar and unfamiliar face-matching task. Furthermore, participants tended to commit more false positive errors in the face matching task when familiar faces were masked, and more false negative errors when unfamiliar faces were masked. Contrary results were observed in a more recent study\cite{kollenda2022influence} on the effect of face masks on familiarity. The authors demonstrate using a face memory task that face familiarity indeed plays a role in face recognition and people tend to remember familiar faces more than unfamiliar faces irrespective of face-masks. Another study\cite{carlaw2022detecting} similarly found effect of familiarity using a recognition without identification (RWI) paradigm but failed to find any difference in RWI effect between face occlusion due to sunglass and surgical masks. In the current study, we explored the effect of face familiarity and face mask on face recognition using a face memory task and also studied the underlying neural mechanism. 

In \cite{noyes2021effect}, the authors studied the effects of masks and sunglasses on familiar and unfamiliar face matching and emotion recognition. Face masks impeded matching performance irrespective of face familiarity however, a different pattern of results was obtained for familiar and unfamiliar faces. While both sunglass and mask produced reduction in performance for unfamiliar faces, occlusion of eyes (sunglasses) did not impair matching performance on the familiar faces. The authors argue that face masks occlude larger face area compared to sunglass and additionally suggest that  our experience with sunglass compared to masks could be one of the possible explanations for difference in performance. Arguing similarly, it is conceivable that after having spent substantial time getting used to face masks, face processing under masked conditions might show lesser performance impairment,however a couple of recent studies failed to show any improvement in masked face perception for face matching\cite{freud2021recognition} and emotion recognition\cite{carbon}. However, in a recent study\cite{carragher2022masked}, the authors demonstrate that with diagnostic face training, impairment in face perception can be improved. In another study using emotion perception\cite{barrick2021mask}, the authors observed a learning effect of face mask exposure and showed that participants used visual cues more effectively with time.  This is contrary to the previous literature\cite{carbon2022reading,freud2021recognition} which argued  that face-processing is an innate ability and unlikely to be altered by environmental changes and argued that face masks ushered a shift in face processing in terms of how people use visual cues for emotion recognition. We wanted to further explore the effect of mask exposure on face familiarity, especially for personally familiar faces, since real life exposure to specific masked faces and not generic masked faces might show some effect of long time mask exposure.

When our present study was first conceived during the second Covid wave in India, face-masks have been in abundant use for more than a year and, people seeing masked-faces would plausibly have had the time to undergo adaptation, albeit possibly temporarily, towards perception of masked-faces. Residing in a secluded residential campus in the suburbs provided us with the unique opportunity to explore the effect of face mask and the associated adaptation effect on face familiarity. During the time of data collection in the current study, only limited number of individuals resided on the campus who were familiar with each other and were continuously seeing other residents with face masks. Thus the campus residents provided the ideal population to test the effect of possible adaptation to face masks. We refer to the possible effect of masked-faces that are continually visualized in one's daily life as "visual experience". 
Our present study, thus, aimed to understand whether such an effect of visual experience is indeed observed empirically, by investigating the differential perception of unmasked and masked faces of differing levels of familiarity. We also explore whether and how visual experience with face-masks might modulate the perception of masked-faces. Thus, we aimed to explore the neural correlates of masked-face perception and the mechanisms by which the same might be affected by face familiarity, and EEG was used for this purpose. 

Neural studies employing EEG have shown the presence of face-sensitive neural signatures in terms of the electrical activity of the brain. We used event related potential (ERP) to explore the effect of mask and familiarity in face processing. We focus on the three early components, P100 , N170 and N250. P100 is an upward-going peak occurring around 100 ms post stimulus display and is associated with early stimulus-driven attentional processes\cite{luck2000event, mangun1995neural,mangun1991modulations}.
The N170 component, an early downward-going EEG potential with a peak approximately around 170 ms post-stimulus onset has been found to be associated with the visual perception of faces. This component is elicited strongly with a larger peak for faces as opposed to non-face objects. The face-sensitive N170 is said to represent the early structural encoding of a face and thereby, act as a marker for face perception \cite{bentin}, \cite{joyce}. The effect of N170 on face familiarity  on the other hand is generally thought to be less consistent\cite{ramon2018familiarity}. Although familiarity effects of N170 are inconsistent across studies, a recent survey paper \cite{caharel2021n170} challenged the standard neurocognitive models and suggested that for personally familiar faces, familiarity with specific facial identities emerges between 150-200 ms in occipito-temporal brain regions.
Another ERP component, N250, a negative-going component occurring approximately around 250 ms post-stimulus onset, has consistently been found to be associated with face-processing, specifically with face familiarity \cite{bentin2000structural,caharel2005familiarity,paller2000electrophysiological,pfutze2002age,tanaka1,tanaka2}. This component has been observed to be more negative (larger peak value) for familiar faces and self-face than for unfamiliar faces. The degree of familiarity has been shown to modulate the N250 component\cite{herzmann2004s,huang2017revisiting}.

Although there have been several behavioral studies exploring effect of face masks on face processing under various condition during covid-19 \cite{carragher, carbon,gori,baron,zochowska,noyes2021effect,prete2022neural}, only two studies reported on the modulation of neural timeline being affected by the face mask. The first study \cite{zochowska} found that attention related components (P100, P300 and LPP) are higher in masked conditions and the second study\cite{prete2022neural} used emotional face stimuli to explore the effect of mask and found significant effect of mask on face related components(N170 and P200). To the best of our knowledge, exploring neural correlates of face identity using face masks has not been reported previously. Additionally, we also explored the role of visual experience on masked face identity. 
In the current study, we attempted to address the underlying neuro-psychological mechanism of face processing when face is occluded by mask and explored the effect of mask on face recognition using personally familiar, famous and unfamiliar face stimuli. Specifically, we address the following questions: (A) How do face-masks affect face processing? (B) How does face-familiarity affect masked face perception? (C) Does visual experience with face-masks play a role in masked face perception? (D) Do the neural correlates of masked face perception differ from those of unmasked face perception?
We used a 2-back test paradigm for our task and used Indian face stimuli consisting of personally familiar, famous (celebrity) and unfamiliar faces.

\section*{Materials and Methods}
\textbf{Ethics statement.} The study was carried out following institutional guidelines. All experimental protocols were approved by the Institute Ethics Committee of the Indian Institute of Science Education and Research (IISER) Kolkata, India. Participants gave written informed consent in accordance with the Declaration of Helsinki.
\\
\subsection*{Stimuli and Display} The data set consisted of 2.67 inch x 4 inch (final viewing size) 8-bit gray-scale images of Indian famous, personally familiar and unfamiliar faces in both unmasked and masked (face-masks on) conditions. There were 20 individuals of each familiarity category (10 female), and 6 face-images (3 unmasked) of each individual. The participants were seated at a distance of approximately 30 inch from the display. Images subtended a visual angle of 5.10 degree x  7.63 degree. The display resolution of the display screen was set at 1366 x 768.
\\
\subsection*{Stimuli Preparation} Images of personally familiar faces were obtained via consensual digital capturing of face photographs of research scholars and residents of IISER Kolkata campus. Images of Indian famous faces licensed under Creative Commons license were obtained from the internet. Images of unfamiliar faces were obtained from an Indian face database \cite{UFdata} licensed under Creative Commons Attribution. All the images were devoid of accessories such as spectacles, sunglasses, hairbands and others. A uniform cloth face-mask was added to each image using a Python code \cite{anwar2020masked} to generate the masked-counterparts of each face image. The face-masks on each individual image were adjusted and edited manually using Photoshop to make each face image appear as natural as possible. Any additional accessories such as earrings were removed at this stage. After adding face-masks, all the images were refined using a software \cite{remini} to match the image resolution across all images. All the images were then cropped to the same size at 283 x 425 pixels (72 pixels/inch) to keep only the face and the hair. Next, the images were converted to gray-scale with an uniform background. Finally, low-level image properties such as mean luminance and contrast were matched across all images using the SHINE toolbox \cite{shine} run under MATLAB environment (ver. 2018b) \cite{MATLAB:2018b}.
\\
\subsection*{Experiment} 41 healthy subjects (ages (years): 21-30, mean: 24.45, std: 2.68, twenty female) participated in the behavioural study; of which 25 subjects (ages (years): 22-30, mean: 24.62, std: 2.69, twelve female) participated in the EEG study based on the same paradigm. Subjects were chosen such that they had spent at least 6 months in campus in order to guarantee them having seen the personally familiar faces in the masked-state. All subjects had normal or corrected-to-normal vision, reported no history of neurological issues, and were naive to the purpose of the study. The experiment consisted of 450 trials split into 15 successive sessions performed in one sitting. The total duration of the experiment was $\sim$ 35 minutes. Each session had an equal representation of all the six possible combinations of familiarity and mask categories. All subjects were right-handed and had used their right hands throughout the experiment to register their responses via mouse-click. 
The sample size was estimated with the MorePower software\cite{campbell2012anova} with estimation conducted for main factor of stimuli('unmasked face', 'masked face') using 2-way repeated measures ANOVA with factors of stimuli and familiarity level ('familiar', 'famous', 'unfamiliar') for estimated effect size $\eta^2=0.25,\alpha=0.05,\beta=0.8$. The computed results indicate a sample size of 24 participants.

\begin{figure}[H]
    \centering
    \includegraphics[scale=0.74]{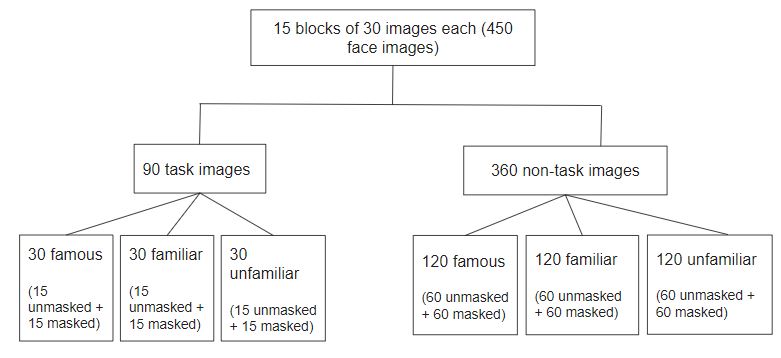}
    \caption{Experiment Design: division of stimuli images of each familiarity and mask category into task and non-task trials}
    \label{fig:exp_design}
\end{figure}

\subsection*{Experimental Paradigm} The experiment was based on a version of the n-back task paradigm \cite{nback}, where n = 2. Within this 2-back task experimental design, subjects were presented with a sequential display of 30 face images (trials) within a session and were required to respond with a left-click on the mouse whenever they believed that they found a repetition, in a 2-back manner, of an individual among the face images displayed. The experiment was conducted for 15 sessions. At the beginning of the experiment, a practice session was run for the subjects to adapt to the requirements of the task design. 20\% of the trials were "task trial" pairs, wherein a true repetition of a stimuli-face occurred, and the rest 80\% were "non-task trial" pairs, wherein the stimuli-face was not repeated in a two-back manner. The sequence of trial presentation for each session was designed in a pseudo-random manner such that each of the six possible combinations of familiarity and mask conditions were uniformly represented in both task-trials and non-task trials, and across each of the 15 sessions.

At the start of each session, instructions for the subjects were listed on the screen for the subjects' convenience. Once ready, the subjects clicked on this screen to start the experiment. Each trial started with a presentation of a fixation cross for 0.8-1 s (uniformly sampled between 0.8 and 1 s), followed by the presentation a face image for 0.1 s, which was followed by a post-stimuli fixation cross for 0.3-0.4 s (uniformly sampled between 0.3 and 0.4 s). Next, for all trials except the first and the second of each session, a response screen was displayed with the text "Response" for a maximum of 1 s, within which the subject had to left-click on the mouse for task trials. Upon a left-click of the mouse, the display screen moved on to the next trial. In case no left-click was registered, the response screen disappeared on the completion of 1 s and the display screen automatically moved on to the next trial. The entire experimental paradigm was coded using Psychtoolbox in MATLAB environment \cite{ptb}.

\begin{figure}[t!]
     \centering
     \begin{subfigure}[b]{1\textwidth}
         \centering
         \includegraphics[width=\textwidth]{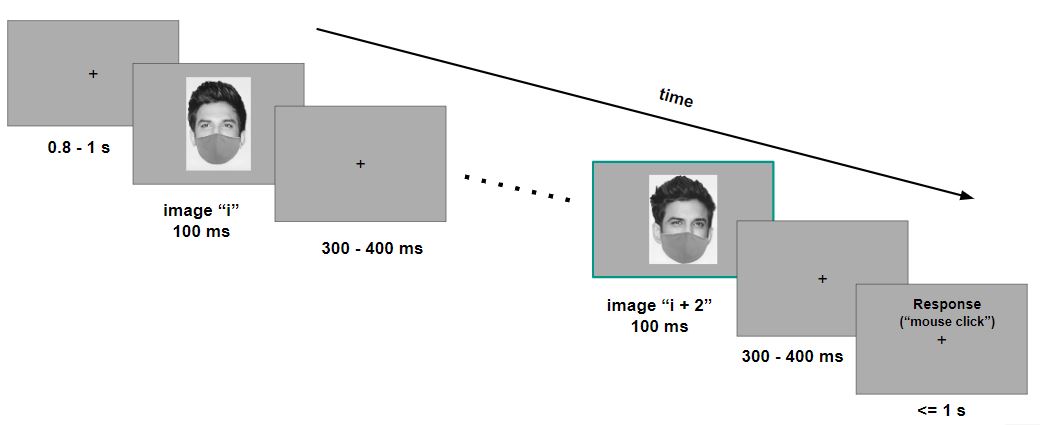}
         \caption{Exemplar of a task trial (in green)}
         \label{fig:task trial}
     \end{subfigure}
     \begin{subfigure}[b]{1\textwidth}
         \centering
         \includegraphics[width=\textwidth]{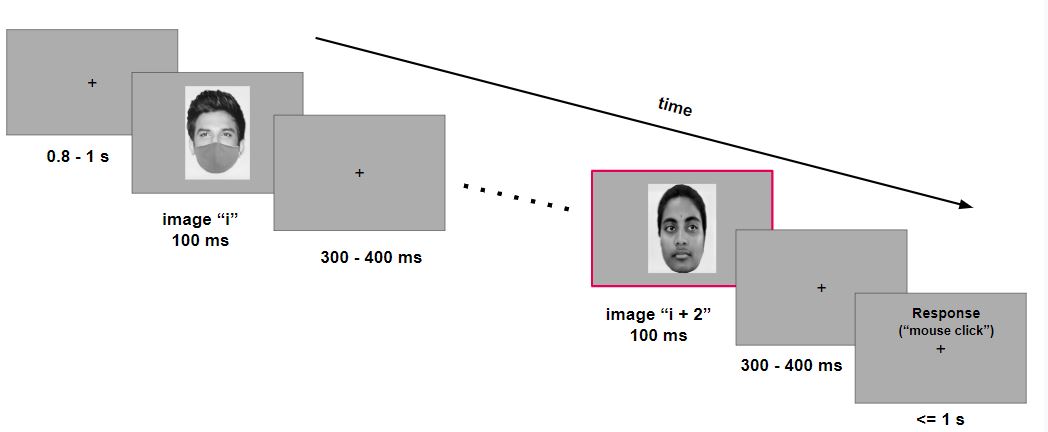}
         \caption{Exemplar of a non-task trial (in red)}
         \label{fig:non task trial}
     \end{subfigure}
     \caption{Experimental Paradigm: 2-back task, showing (a) a task trial, and (b) a non-task trial}
        \label{fig:paradigm}
\end{figure}

At the end of all 15 sessions, subjects were required to complete a post-experiment familiarity rating task wherein subjects were presented with images of the personally familiar and famous stimuli individuals and were asked to indicate on a scale of 1-10 how familiar they were, at the time of participation, to the personally familiar and famous stimuli individuals. A rating of 10 meant that the subject was certainly familiar while a rating of 1 indicated that the subject was certainly unfamiliar with the stimuli individual in question. For each subject, those trials that consisted of personally familiar and famous stimuli individuals who were rated a score lower than 6 were removed from further analysis. For subjects finally considered for analysis, the average percentage of personally familiar and famous stimuli excluded from analysis on the basis of this familiarity-rating task were 27.14\% and 2\%, respectively.
\\
\subsubsection*{Behavorial Response Parameters} By virtue of the experiment being a two-back task paradigm, there were four possible behavioural responses that could be registered by a subject on each experimental trial: a true-positive (TP) response, wherein a subject clicks on a task trial; a true-negative (TN) response, in which a subject does not click on a non-task trial; a false positive (FP) response (or, a false alarm), wherein a subject clicks on a non-task trial; and a false negative (FN) reponse (or, a miss), in which a subject does not click on a task trial.

Two types of behavioural responses were collected as a measure of behavioural performance in the experimental task: 

Performance Accuracy (PA): It is a measure of how well a subject performed in the above-mentioned experimental task, and is defined as
    \begin{equation}\label{eq:1}
        PA = \frac{TP}{TP+FN}
    \end{equation}
where, TP and FN are the total number of true positive and false negative responses, respectively, registered by the subject over all 15 experimental sessions.
    
 Reaction Time (RT): It is calculated as the time between the onset of an image and the moment when the subject registers a click. This was measured using functions available in Psychtoolbox, MATLAB (\cite{ptb}). Although RTs were measured every time a subject registered a left-click on the mouse (that is, for TP as well FP responses), only those RTs corresponding to the TP responses were analyzed for the purposes of this study because RT as a measurable behavioral parameter was intended to serve as a measure of how well a subject performs in the above-mentioned experimental task.
\\        


\subsection* {Neural Data Acquisition and Preprocessing.}
EEG activity was measured using a 64 channel active shielded electrodes mounted on an EEG cap which follows the International 10-20 system. EEG data was collected at a sampling rate of 512 Hz. Trials were band-pass filtered, to filter out those neural signals with frequencies lower than 0.1 Hz or greater than 40 Hz. The EEG signals, being a continuous measure of brain activity, were epoched with respect to stimulus onset, and were referenced using average referencing. These preprocessed EEG epochs were time-locked to stimulus onset and consisted of a 200 ms pre-stimulus baseline and a 400 ms post-stimulus interval, and are referred to as EEG trials. Trials affected by eye-movements, blinking or motor movements were rejected using a semi-automatic approach with initial epoch rejection using EEGLAB toolbox \cite{eeglab} followed by manual rejection of noisy trials.

\subsection*{Data Analysis}
The experiment intended to test the effect of two experimental factors, namely, familiarity and face-mask, on the performance of human subjects in a 2-back task. In this context, the experimental conditions were decided as follows:
For the familiarity factor, the three experimental conditions were famous, familiar, and unfamiliar (control).
For the mask factor, the two experimental conditions were masked (with face-mask), and unmasked (without face-mask, control).
The behavioural data was collected in the form of performance accuracy and reaction time, whereas the neural data were in the form of EEG signals which were analyzed in the time-domain. Thus, the dependent variables were performance accuracy and reaction time, for the behavioural data, and EEG potentials for the neural data. The independent variables, in both cases, were the familiarity and mask conditions. All the analyses were run in MATLAB.

\subsubsection*{Behavioural Data Analysis}
The performance accuracy for each subject was calculated as shown in Equation \ref{eq:1}. In such an experimental paradigm, wherein each task trial has only two possible outcomes (a mouse-click, or no mouse-click), a chance-level performance accuracy (that is, if the subject were to randomly click on the mouse button for each trial) would be 50\%. Subjects with chance-level or lower performance accuracy were not considered for further analysis. Thus, 5 out of 41 subjects, having mean performance accuracy lower than 50\% across all trials of unmasked familiar and unmasked famous faces, were removed from further analysis. For the remaining 36 subjects, a two-way repeated measures ANOVA \cite{rmanova2} analysis was run to check for the effects of the two factors of familiarity (with 3 levels) and mask-condition (with 2 levels) on each of the two behavioral response parameters . In case a significant main effect was observed in the ANOVA, post-hoc analyses were done using Tukey's HSD test. Two-way repeated measures ANOVA was also run to test for effects of the independent variables on the proportion of TP, FN, FP, and TN responses.
\\

\subsubsection*{Neural Data Analysis}

Neural analysis focused on ERP components of P100, N170, and N250. For the N170 and N250 ERP components, the left-brain and right-brain electrode clusters were chosen as TP7, P7, P9, PO3, and PO7, and TP8, P8, P10, PO4, and PO8, respectively (\cite{tanaka1}, \cite{tanaka2}, \cite{zochowska}). whereas, for the P100 ERP components, the electrode clusters were chosen as O1, O2, PO3 and PO4 (Fig. \ref{fig:left_right_P100}), respectively (\cite{zochowska}). Peak based amplitude was considered for P100 and N170 whereas mean amplitude over 230-320 ms was considered for N250 component. 5 subjects with noisy EEG signals were excluded from analysis. For each subject, all (task as well as non-task) EEG trials were segregated into the 6 conditions (combination of the factors of familiarity and masking), and for each such condition the EEG trials were averaged to give subject-wise ERPs. The mean of each such subject-averaged ERPs were taken to generate the grand-averaged ERPs for each condition. Those task trials of the 2-back task paradigm on which the subject has performed correctly (henceforth referred to as correct task-trials) were analyzed in terms of their corresponding ERP amplitudes. The computed ERP components were tested for significance using one way and two way repeated measure ANOVA models. Using two way ANOVA with mask and familiarity as two conditions for ERP analysis however resulted in significant interaction effect \cite{multivariate} for N170 and N250 thereby rendering the statistical inference not interpretable. Hence for meaningful interpretation, we use five one-way repeated measure ANOVA by fixing each of the five levels (Unmasked, Masked, Familiar, Famous, and Unfamiliar) of the two experimental fsctors. If significant main effect was present in the one way ANOVA models, follow up analysis using 2 way repeated measure ANOVA with hemisphere as additional factor was used.



\section*{Results}

\subsection*{Behavioural Findings}

A two-way repeated measures ANOVA showed significant main effects of both the factors of familiarity and face-mask on both the measures of performance accuracy as well as reaction time as summarized in table \ref{tab:anova_behaviour}.

The two-way ANOVA revealed significant main effects of familiarity (F(2,70) = 10.07, $p$<0.001) as well as mask-condition (F(1,35) = 12.44, $p$<0.01) on performance accuracy. Further, ANOVA also revealed significant main effects of familiarity (F(2,70) = 12.62, $p$<0.0001) and mask-condition (F(1,35) = 10.73, $p$<0.01) on reaction time. Notably, the interaction effect in both cases were non-significant. 

ANOVA tests were required to be followed by multiple comparison tests because there were more than two levels of the familiarity factor. A post-hoc Tukey's test was chosen as the multiple comparison test to identify which of the means among the experimental conditions are significantly different from each other.


\subsubsection*{Performance Accuracy} 

\begin{figure}[H]
\centering
\includegraphics[scale=0.7]{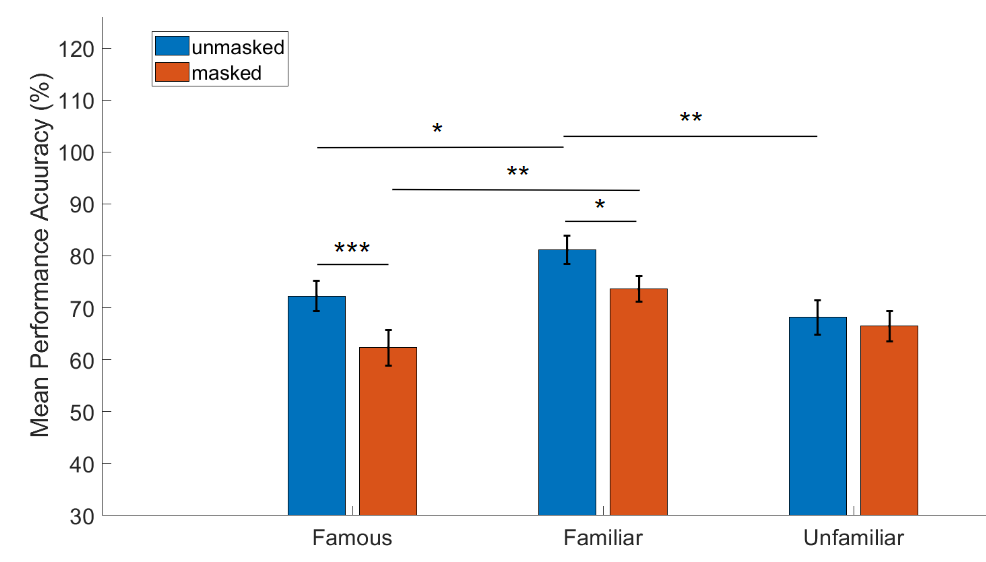}
\caption{Effect of familiarity and face-mask on performance accuracy: results of two-way repeated measures ANOVA followed by Tukey's multiple comparison test (N = 36, error bars indicate standard error of mean, \textit{*** $p$<0.001, ** $p$<0.01, * $p$<0.05})}\label{fig:PA_all}
\index{figures}
\end{figure}
The two-way repeated measures ANOVA on the measure of performance accuracy was followed by post-hoc Tukey's test, the results of which are described as follows: mean performance accuracy for unmasked famous faces was significantly higher than that obtained for its masked counterpart ($p$<0.001, CI=[0.04, 0.15], BF=39.85), whereas level of significance and corresponding Bayes factor was comparatively less when compared between unmasked familiar and masked familiar faces ($p$=0.02, CI=[0.01, 0.14], BF=2.14). Mean performance accuracy was also found to be significantly lower for unmasked famous than for unmasked familiar ($p$=0.02, CI=[-0.16,-0.01], BF=6.02), however this difference was stronger while comparing masked famous and masked familiar faces ($p$=0.003, CI=[-0.19,-0.04], BF=30.92). Mean performance accuracy was also found to be significantly higher for unmasked familiar faces than for unmasked unfamiliar faces ($p$=0.004, CI=[0.04,0.22], BF=23.6).



\subsubsection*{Reaction Time (RT)} 

\begin{figure}[H]
\centering
\includegraphics[scale=0.75]{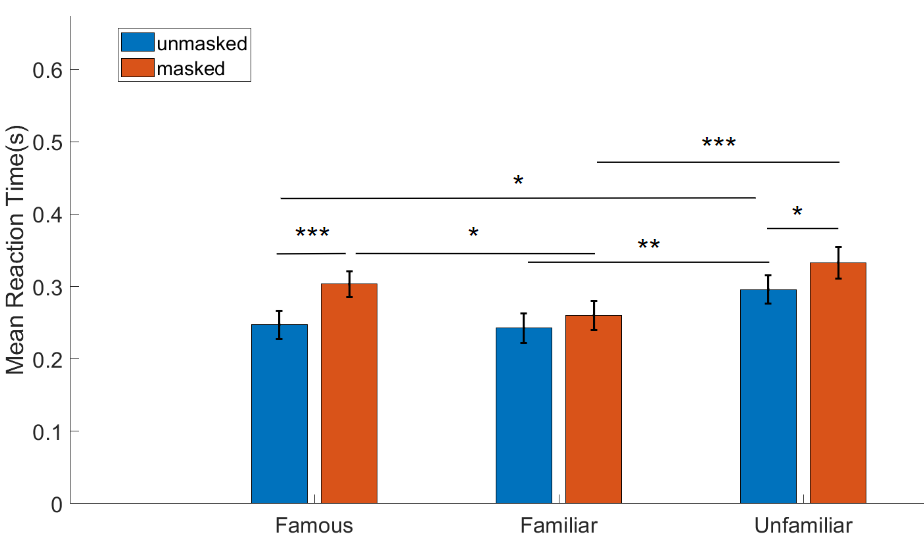}
\caption{Effect of familiarity and face-mask on reaction time: results of two-way repeated measures ANOVA followed by Tukey's multiple comparison test (N = 36, error bars indicate standard error of mean, \textit{*** $p$<0.001, ** $p$<0.01, * $p$<0.05})}\label{fig:RT_all}
\index{figures}
\end{figure}
The two-way repeated measures ANOVA on the measure of RT was followed by post-hoc Tukey's test, the results of which are described as follows: mean RT for unmasked famous faces was significantly lower than that obtained for its masked counterpart ($p$<0.001, CI=[-0.09,-0.03], BF=43.42), and mean RT for unmasked and masked familiar faces did not differ significantly. Mean RT for unmasked unfamiliar was significantly lower than that obtained for masked unfamiliar faces ($p$=0.01, CI=[-0.07,-0.01], BF=3.61). Mean RT was also found to be significantly lower for unmasked familiar than for unmasked unfamiliar ($p$=0.003, CI=[-0.09,-0.02], BF=30.66). Mean RT was also lower for unmasked famous faces than for unmasked unfamiliar faces ($p$=0.01, CI=[-0.09,-0.01], BF=8.06). Mean RT was found to be significantly higher for masked famous faces than for masked familiar faces ($p$=0.04, CI=[0.002,0.08], BF=3.13). There was a strong significant increase in mean RT for masked unfamiliar faces when compared to masked familiar faces ($p$<0.001, CI=[-0.11,-0.04], BF=991.83).



\subsubsection*{False Positive (FP) Response}

\begin{figure}[H]
\centering
\includegraphics[scale=0.6]{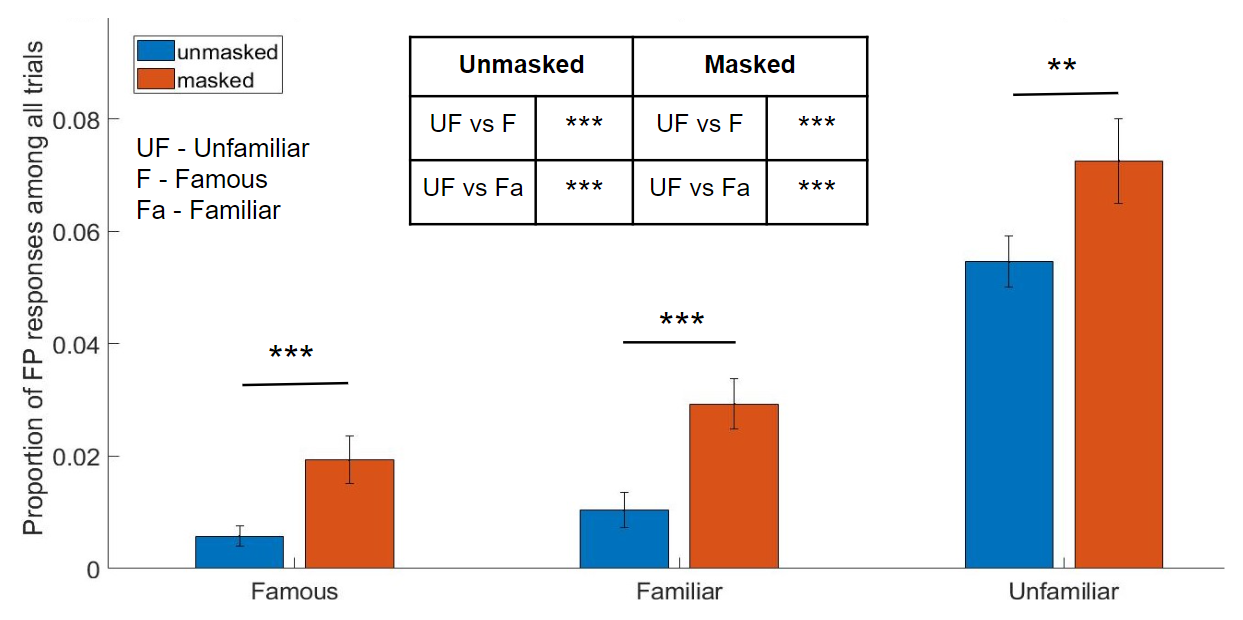}
\caption{Proportion of false positive responses (error bars indicate standard error of mean, \textit{** $p$<0.01, *** $p$<0.001})}\label{fig:FP_all}
\index{figures}
\end{figure}

The two-way ANOVA revealed significant main effects of familiarity (F(2,70) = 93.77, $p$<0.0001) as well as mask-condition (F(1,35) = 33, $p$<0.0001) on the proportion of FP responses. Notably, the interaction effect was non-significant. 

The two-way repeated measures ANOVA on the proportion of FP responses was followed by post-hoc Tukey's test, the results of which are described as follows: FP responses for unmasked famous faces was significantly lower than that obtained for its masked counterpart ($p$<0.0001, CI=[-0.02,-0.01], BF=133.68). The same trend (that of FP proportion being lower in unmasked condition as opposed to masked) was observed for familiar ($p$<0.0001, CI=[-0.03,-0.01], BF=212.15), as well as unfamiliar faces ($p$=0.006, CI=[-0.03,-0.01], BF=6.25). The proportion of FP responses was also found to be significantly lower for unmasked familiar than for unmasked unfamiliar ($p$<0.0001, CI=[-0.05,-0.03], BF=\SI{2.67e10}). FP responses was also lower for unmasked famous faces than for unmasked unfamiliar faces ($p$<0.0001, CI=[-0.06,-0.04], BF=\SI{1.52e10}). FP response proportion was found to be significantly lower for masked famous faces than for masked unfamiliar faces ($p$<0.0001, CI=[-0.07,-0.04], BF=\SI{8.05e6}). FP response proportion was also found to be significantly lower for masked familiar faces as compared to masked unfamiliar faces ($p$<0.0001, CI=[-0.06,-0.03], BF=\SI{1.48e4}).

\subsubsection*{Behavioral Results Summary}
There was an overall significant effect of mask as well as familiarity. Familiar face recognition gave the best performance followed by famous face  irrespective of masked condition and reverse trend was observed for reaction time. The masked condition resulted in significantly decreasing the performance accuracy and increasing the reaction time for famous face recognition when compared with unmasked condition. There was no significant change in terms of performance in the unfamiliar face identity for masked and unmasked condition however reaction time for unmasked unfamiliar faces was significant when compared to the masked counterpart. In the context of familiar face recognition, although masked condition showed slightly reduced performance accuracy but the level of significance for masked condition was less compared to famous face stimuli and there was no significant difference in reaction time between masked and unmasked familiar face recognition. Masked unfamiliar face category resulted in highest false positive errors pointing to the inherent difficulty in processing unknown faces compared to familiar faces. These results are indicative of the possible role of visual experience in identifying frequently viewed masked faces.

\subsection*{Neural Findings}


The event-related potentials of the three early components (P100, N170 and N250) was analysed for true positive responses. For the P100 ERP components, the electrode cluster was chosen as O1, O2, PO3, and PO4 electrodes (\cite{zochowska}). For N170 and N250, the left-brain and right-brain electrode clusters were chosen as TP7, P7, P9, PO3, and PO7, and TP8, P8, P10, PO4, and PO8, respectively. Furthermore, ERP results obtained for the left and right hemisphere combined is also shown (Fig. \ref{fig:erp_N170_N250}) (\cite{tanaka1}, \cite{tanaka2}, \cite{zochowska}.

\subsubsection*{P100}

\begin{figure}
    \centering
    \includegraphics[scale=0.6]{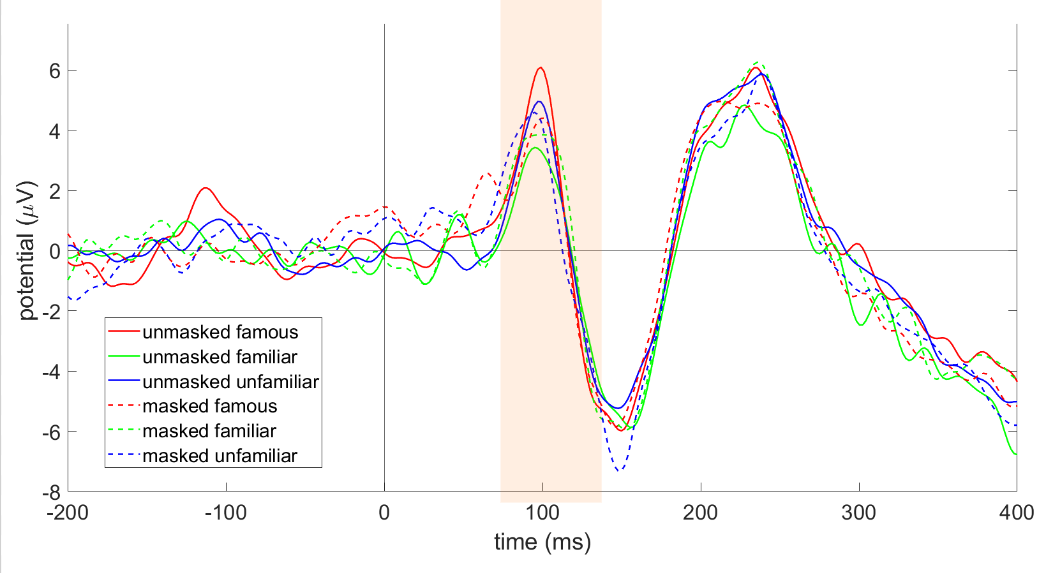}
    \caption{P100: Grand-averaged ERP across all electrodes in the cluster (O1, O2, PO3, PO4) (N = 20)}
    \label{fig:left_right_P100}
\end{figure}

P100 component displayed no significant main effect of mask ($p=.28$) and familiarity ($p=0.78$) condition with insignificant interaction effect in a two-way repeated measure ANOVA model. The result is also consistent when looked separately at each hemisphere electrode-cluster. 

\subsubsection*{N170}

\begin{figure}
    \centering
    \begin{subfigure}[b]{0.45\textwidth}
    \includegraphics[scale=0.4]{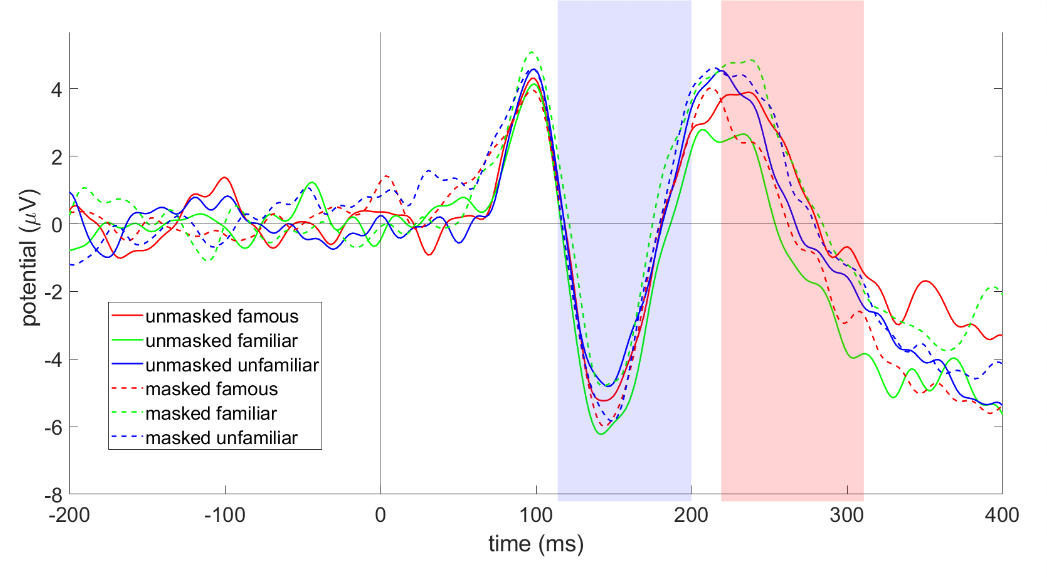}
    \caption{Right hemisphere cluster}
    \label{fig:N170_N250_right}
    \end{subfigure}
     \begin{subfigure}[b]{0.45\textwidth}
    \includegraphics[scale=0.4]{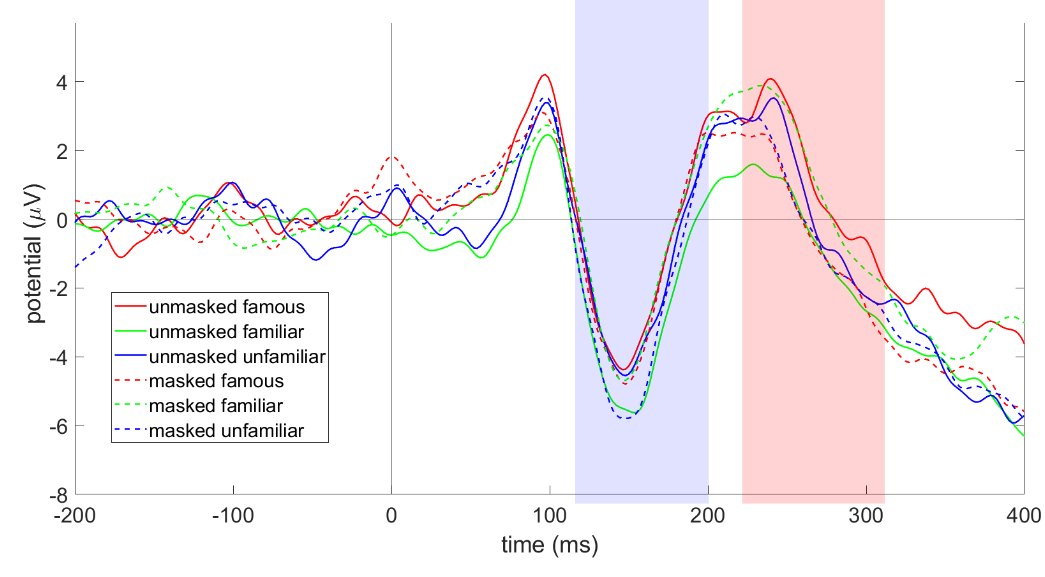}
    \caption{Left hemisphere cluster}
    \label{fig:N170_N250_left}
    \end{subfigure}
     \begin{subfigure}[b]{1\textwidth}
     \centering
      \includegraphics[scale=0.6]{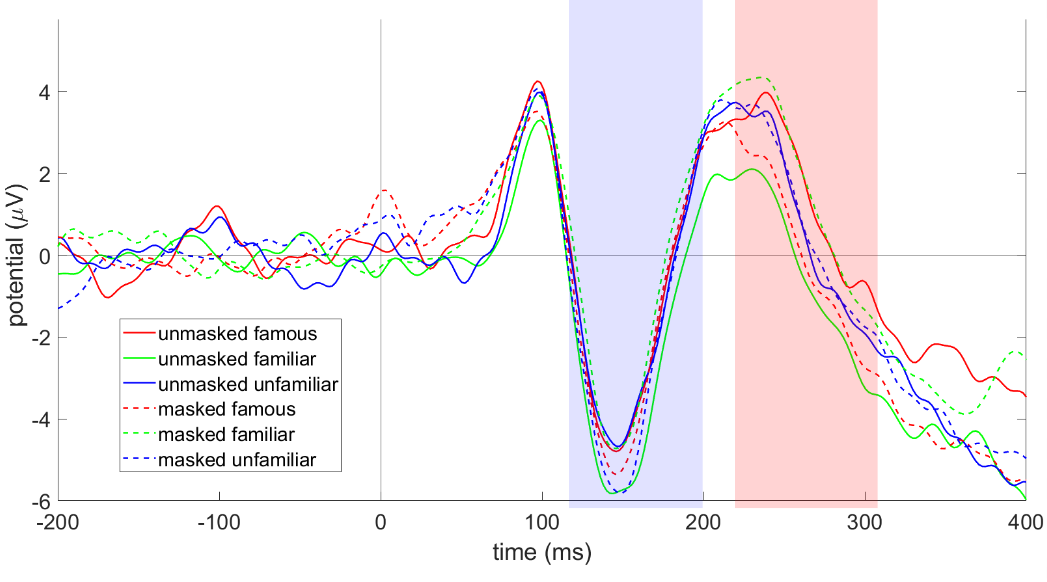}
    \caption{Right and Left clusters combined}
    \label{fig:N170_N250_whole}
    \end{subfigure}
    \caption{ERP plots of electrode clusters for N170 and N250 ERP components (blue band: N170, red band: N250), grand averaged over all subjects (N=20), showing (a) Right Hemisphere electrode cluster (TP8, P8, P10, PO4, PO8), (b) Left Hemisphere electrode cluster (TP7, P7, P9, PO3, PO7), and (c) Right and Left electrode clusters combined (TP7, TP8, P7, P8, P9, P10, PO3, PO4, PO7, PO8)}
    \label{fig:erp_N170_N250}
\end{figure}

The one-way repeated measures ANOVA resulted in a significant difference in the N170 response between unmasked familiar and masked familiar conditions (F(1,17)=4.86, $p$=0.04). Using a two-way repeated measures ANOVA framework, with mask-condition and hemisphere as the two factors (Table \ref{tab:anova_N170_2_hemi_familiar}), unmasked familiar and masked familiar N170 response resulted in a significant main effect of mask (F(1,17)=4.75, $p$=0.044) but there was no significant effect of hemisphere. Interaction effect was also found to be non-significant. No significant effects were found on post-hoc analysis (weakly significant effects were found for left-unmasked vs left-masked (p=0.055), and for right-unmasked vs right-masked (p=0.07)).

\subsubsection*{N250}

The one-way repeated measures ANOVA resulted in a significant difference in the N250 response between unmasked familiar and masked familiar conditions (F(1,17)=4.58, $p$=0.02), as also between unmasked famous and masked famous conditions (F(1,17)=6.55, $p$=0.047). Using a two-way repeated measures ANOVA framework including hemisphere  with mask-condition and hemisphere as the two factors (Tables \ref{tab:anova_N250_2_hemi_familiar},\ref{tab:anova_N250_2_hemi_famous}, respectively) gave the results as described below.

The two-way repeated measures ANOVA on unmasked familiar and masked familiar N250 response resulted in a significant main effect of mask (F(1,17)=6.55, $p$=0.02) but there was no significant effect of hemisphere. Interaction effect was also found to be non-significant. On running a post-hoc Tukey test, the N250 for unmasked familiar condition was found to be significantly lower than that obtained for masked familiar condition as observed in the right-hemisphere electrode cluster ($p$=0.01, CI=[-4.4,-0.6], BF=4.06). No other significant effects were found on post-hoc analysis.

The two-way repeated measures ANOVA on unmasked famous and masked famous N250 response resulted in a significant main effect of mask (F(1,17)=4.58, $p$=0.047) but there was no significant effect of hemisphere. Interaction effect was also found to be non-significant. On running a post-hoc Tukey test, the N250 potential for unmasked famous condition was found to be significantly greater than that obtained for masked famous condition as observed in the right-hemisphere electrode cluster ($p$=0.048, CI=[0.01,3.4], BF=1.48). No other significant effects were found on post-hoc analysis.

\subsubsection*{Neural Results Summary}
The ERP analyses demonstrated that there is no difference between mask and familiarity conditions for the P100 component. However, there was a significant difference found in N170 between unmasked and masked familiar face recognition. Both famous and familiar face recognition differed between masked and unmasked condition in the N250 component.
In summary, mask effect was evident for personally familiar faces as early as N170 whereas for famous faces, the mask effect was prominent after 250 ms. 

\section*{Discussion}

Face processing deserves a special place in our life in terms of the social information it provides including age, gender, emotion, identity and others. However the advent of COVID-19 pandemic and use of face masks have severely affected the way we socially interact with each other. In this study, we explored the effect of mask and familiarity on face recognition using a 2-back task and analysed the related neural correlates.

\subsection*{Effect of familiarity and face mask on face recognition}
Face masks used generously during the pandemic have significantly affected our ability to process and recognize face. Previous behavioral studies in the past 2 years have shown that face masks had a detrimental effect on face processing and recognition\cite{carragher,prete2022neural,noyes2021effect,carbon,gori,zochowska,kollenda2022influence}. Effect of face mask has also been evident during emotion recognition\cite{gori,carbon,carbon2022reading}. Our study shows that in unmasked condition, recognition was best for personally familiar faces followed by famous faces while participants performed the worst when encountering an unfamiliar face. This is in concurrence with familiar face recognition literature \cite{ramon2018familiarity,carbon2008famous}. In the masked condition, performance decreased for all categories following the same trend shown in recent studies\cite{carragher,zochowska,noyes2021effect}, however, the masking effect was most prominent for famous faces (decrease in performance accuracy by 13.39\%) followed by personally familiar faces(5.82\%). There was no significant effect of face masks on performance accuracy for unfamiliar faces. The results are consistent with our reaction time analyses as well for famous face recognition while there was no significant difference in personally familiar face recognition in the masked condition while unfamiliar face recognition shows slight difference in reaction time. The unfamiliar faces also showed the most false positive response compared to personally familiar and famous faces. Our results show significant effect of both familiarity and mask on face recognition and demonstrate the detrimental effect of face mask on recognition irrespective of familiarity (no significant interaction between familiarity and masks). We discuss the effect of mask on famous and personally familiar face recognition separately in the following sections.

\subsubsection*{Effect of mask on famous and unfamiliar face recognition}
Most of the recent face familiarity studies using face masks use famous faces\cite{carragher, noyes2021effect} and face matching tasks. Our results show that the impact of face masks was most prominent during famous face recognition and convincingly shows the effect of familiarity in both masked and unmasked condition while previous study\cite{carragher} shows the opposite trend. However recent studies\cite{kollenda2022influence,carlaw2022detecting} shows effect on familiarity on masked face recognition which is in line with our results. A previous study\cite{noyes2021effect} also found deterioration in performance for unfamiliar face compared to familiar face matching in cross-experimental comparisons. In \cite{carragher}, the authors interpreted the insignificant effect of familiarity as inconsequential because of different baseline performances for the two conditions. In the experiments citing no effect of familiarity in masked condition, the task was relatively simple with \cite{carragher} using a face matching task where both images were shown simultaneously. However in studies showing familiarity effect using face mask including the current study, the task was more challenging. We use a 2-back test while \cite{carlaw2022detecting} uses a recognition without identification paradigm and in \cite{kollenda2022influence}, short-term memory for famous and unfamiliar identities in masked conditions were used. Participants identify the two faces within the same trial in matching tasks whereas in face memory tasks including ours, the participants use stored representation of faces shown previously to compare with the test image. This difference in paradigm might potentially contribute to the difference in results wherein less challenging face recognition tasks does not show any familiarity effect using face masks. The face recognition performance using within design paradigm could contribute to discrepancy in results from a memory task which uses a between design paradigm \cite{fysh2018individual}.

\subsection*{Role of visual experience for personally familiar faces}
One of the many unprecedented changes that COVID-19 pandemic enforced on us is the use of face masks and recent studies have demonstrated that a general deterioration in face processing and recognition can be attributed to the use of face masks\cite{carragher, prete2022neural, carbon, noyes2021effect}. However, whether the effect of these forced exposure to masked faces in daily life has contributed to a significant change in face perception and recognition remains relatively unknown. Two longitudinal studies\cite{carbon2022reading,freud2021recognition} conducted over the last two years using famous faces have failed to show any improvement in performance  and shows persistent deficits in recognizing masked faces. Individual experience with masked faces was not shown to be correlated with the mask effect and these results points to the mature face processing system being not amenable to changes following prolonged exposure to masks. However, one study\cite{barrick2021mask}
demonstrated effect of visual experience in how people process emotional faces by showing that with prolonged experience in face masks, participants used more global visual cues for emotion processing. In a separate study \cite{carragher2022masked}, it was shown that with diagnostic training, it was possible to improve face processing under mask condition. In the current study we show evidence of visual experience using face masks in face recognition for personally familiar faces when compared with famous faces. Our results show that the difference in performance for personally familiar faces with and without face masks was much lower when compared with difference in famous faces. This conclusion is also validated by reaction time analysis where there was no significant difference between masked and unmasked personally familiar face recognition.
Our participants residing in an isolated campus was forced to encounter personally familiar faces for at least 6 months before the experiments were conducted. Thus our participants unlike the participants in the longitudinal studies were not only used to seeing masked faces, but they were forced to view masked faces of the stimuli in real life for a prolonged period prior to the experiment. Real-life interaction with masked personally familiar faces possibly have a role to play in relatively improved performance under masked condition thus alluding to the role of visual experience. Although it can be argued that the enhanced performance for personally familiar faces (with and without face masks) can be attributed simply to the superior performance of personally familiar faces compared to famous faces as evidenced in face familiarity studies\cite{ramon2018familiarity,carbon2008famous}, however, the neural results show a different trend for personally familiar faces with masks. The neural correlates of masked familiar faces also differ significantly from famous faces further pointing to the possible role of visual experience. Future studies using large sample size of personally familiar faces could perhaps clarify the role of visual experience of face masks on face recognition capability and also shed light on whether the effect of visual experience is sustained for a longer time period.

\subsection*{Neural Correlates of face familiarity and face masks}
In the current study we explore the effect of face mask on famous, personally familiar and unfamiliar face processing along with their underlying neural correlates. We have analysed three related ERP components (P100, N170 and N250) and shown that significant difference exists between masked and unmasked personally familiar face recognition in both N170 and N250. Masking effect was evident for famous faces in only N250 suggesting that masking effect for personally familiar faces starts as early as 170 ms while for famous faces, the onset is delayed. There have been two neural studies\cite{zochowska,prete2022neural} investigating  the effect of face masks out of which one study\cite{prete2022neural} explored the effect of face masks on emotion processing and found N170 and P2 were affected by face masks. The other study\cite{zochowska} which is relevant to our work uses self-image, personally familiar and unfamiliar images to explore the effect of face masks.The participants performed a purely attentional task of simply detecting presented face images. In this study, the authors reported that face masks enhanced the early(P100) and late(LPP, P300) attentional components while no such changes were evident for N170. It also showed slower latency for early ERP components (P100,N170) using masked faces. Overall the effects related to surgical-like masks were similar for all faces which is in line with \cite{carragher}. Our study however shows different results and we find familiarity effect using face masks and it is more prominent for personally familiar faces starting from 170 ms onwards. As suggested previously, the difference in results can possibly be attributed to the difference in experimental task since our study uses a 2-back paradigm which is a more challenging task using working memory than a simple detection task used in\cite{zochowska}. 

Whether N170 plays a role in face familiarity is a pending issue in face recognition and there is evidence supporting both views (see \cite{ramon2018familiarity} for review). Since majority of studies report consistent effect of familiarity in the later ERP components like N250\cite{schweinberger2002event,wiese2019later,wiese2021familiarity,tanaka2006activation,bentin2000structural,eimer2011face}, it is generally assumed that N170 reflects the perceptual awareness stage in face recognition akin to "structural encoding" stage in Bruce and Young's standard model of face recognition\cite{bruce1986understanding}, and the later ERP components play a more prominent role in face identification using stored memory representations\cite{bentin2000structural,eimer2000event,eimer2000effects,eimer2011face}. However, there has been evidence in literature for earlier processing of face familiarity
and the effect is more prominent for personally familiar faces\cite{ramon2018familiarity,caharel2002erps,caharel2005familiarity,caharel2006effects,caharel2021n170,barragan2015neural,huang2017revisiting}. In a recent review paper \cite{caharel2021n170}, the authors studied the role of N170 in several face familiarity studies and concluded that face familiarity enhances the N170  component and the effect is stronger for personally familiar faces. The authors argue that onset of familiarity with specific face identities starts as early as 150-200 ms in occipito-temporal areas shortly following emergence of face selectivity which emerges right before familiarity effect. This observation is in contrast with the standard neuro-cognitive models which distinguishes between perceptual and memory processing in human face recognition.
Our results seems to be consistent with the authors in \cite{caharel2021n170} and shows a significant difference between masked and unmasked faces for personally familiar faces for the N170 component. N170 is visibly enhanced even for unmasked familiar faces when compared with other unmasked faces (see Figure 7), however the difference does not reach significance. The N170 peak is enhanced for masked condition in case of famous and unfamiliar faces which is consistent with \cite{prete2022neural} however the difference in not significant. Interestingly, the N170 peak amplitude decreases in the masked condition for personally familiar faces and the same trend is observed for the N250 component as well. The difference between masked and unmasked faces becomes more prominent in the N250 component and there is significant difference for both famous and personally familiar faces. The N250 is more enhanced for masked famous face and could possibly be attributed to more effort required due to the mask and the effect is in line with studies reporting other structural manipulations of the face\cite{civile2012face,colombatto2017effects}. N250 is assumed to be a reliable marker of face familiarity and has been shown to reflect memory activation\cite{ramon2018familiarity}. Thus, it is not surprising that familiarity plays a role in distinguishing between masked and unmasked face stimuli. However, what is surprising is for personally familiar faces, N250 amplitude decreases and this trend is opposite to the trend observed in famous faces. However, this decrease in amplitude for personally familiar face is consistent for both N170 and N250 and we speculate that the decrease can be potentially attributed to visual experience. However, further explorations are needed to corroborate this possible link, but our study provides one of the earliest evidence of cortical alteration for known face recognition components due to prolonged exposure to face masks.
Further studies are needed to investigate the role of mask and familiarity but the current results show convincingly that our cortical activity is transforming after being exposed to masks for a long period of time. Systematic future study of the transformed neural activity can potentially enhance our current understanding of face familiarity effect in general.

\section*{Conclusion}

The current study explores effect of face mask and familiarity on face recognition along with the underlying neural correlates facilitating the recognition. Our results suggest that face masks can be a contributing factor for face recognition impairment and recognition is impeded by masks contributing to the decrease in performance.  Our results further suggests possible role of N170 and N250 in modulating neural signals under masked condition thus altering neural representation during face recognition. 
Although our results show a comparable detrimental effect of face masks on personally familiar and unfamiliar face recognition yet the underlying cause behind the similar effect is possibly completely different. We hypothesize that the reduction of impairment level using face masks for  personally familiar faces (when compared to famous faces) can be probably attributed to visual experience while for unfamiliar face recognition, the masking effect was not very significant because of the inherent difficulty of remembering unknown faces matching the target image in a face memory task. The large number of false positives for unfamiliar face (with and without masks) can be cited to support our hypothesis. Hence, irrespective of masks, unfamiliar face recognition using visual memory is challenging whereas similar bias is not evident in famous and personally familiar faces. Our results demonstrate that people remember personally familiar faces better even while using face masks alluding to the role of visual experience while the impairment due to mask is more prominent for famous faces.

\section*{Supplementary Information}

\subsection{Behavioural Findings}
\begin{table}[hbt!]
\centering
\begin{adjustbox}{width=0.6\textwidth}
\begin{tabular}{|c|c|c|}
\hline
\hline
\multicolumn{3}{|c|}{\textbf{Dependent Variable: Performance Accuracy}}\\
\hline
\hline
\textbf{Factor} & \textbf{F statistic} & \textbf{$p$-value}\\
\hline
Familiarity & 10.07 & \cellcolor{yellow!50}<0.001\\
\hline
Mask & 12.44 & \cellcolor{yellow!50}0.0012\\
\hline
Interaction & 2.12 & 0.1273\\
\hline
\hline
\multicolumn{3}{|c|}{\textbf{Dependent Variable: Reaction Time}}\\
\hline
\hline
\textbf{Factor} & \textbf{F statistic} & \textbf{$p$-value}\\
\hline
Familiarity & 12.62 & \cellcolor{yellow!50}<0.0001\\
\hline
Mask & 10.73 & \cellcolor{yellow!50}0.0024\\
\hline
Interaction & 2.16 & 0.1225\\
\hline
\end{tabular}
\end{adjustbox}
\caption{Two-way repeated measures ANOVA on performance accuracy and reaction time}\label{tab:anova_behaviour}
\index{tables}
\end{table}

\begin{table}[H]
\centering
\begin{adjustbox}{width=\textwidth}
\begin{tabular}{|m{40mm}|m{40mm}|m{25mm}|m{25mm}|}
\hline
\hline
\multicolumn{4}{|c|}{\textbf{Dependent Variable: Performance Accuracy}}\\
\hline
\hline
\hfil \textbf{Group 1} & \hfil \textbf{Group 2} & \hfil \textbf{$p$-value} & \hfil \textbf{95\% C. I.}\\
\hline
\hfil Unmasked Famous & \hfil Masked Famous & \hfil \cellcolor{yellow!50}<0.001  & \hfil [0.04, 0.15]\\
\hline
\hfil Unmasked Familiar & \hfil Masked Familiar & \hfil \cellcolor{yellow!50}0.02 & \hfil [0.01, 0.14]\\
\hline
\hfil Unmasked Unfamiliar & \hfil Masked Unfamiliar & \hfil 0.59 & \hfil [-0.05, 0.08]\\
\hline
\hfil Unmasked Famous & \hfil Unmasked Familiar & \hfil \cellcolor{yellow!50}0.02 & \hfil [-0.16, -0.01]\\
\hline
\hfil Unmasked Famous & \hfil Unmasked Unfamiliar & \hfil 0.48 & \hfil [-0.04, 0.13]\\
\hline
\hfil Unmasked Familiar & \hfil Unmasked Unfamiliar & \hfil \cellcolor{yellow!50}0.004 & \hfil [0.04, 0.22]\\
\hline
\hfil Masked Famous & \hfil Masked Familiar & \hfil \cellcolor{yellow!50}0.003 & \hfil [-0.19, -0.04]\\
\hline
\hfil Masked Famous & \hfil Masked Unfamiliar & \hfil 0.4 & \hfil [-0.12, 0.04]\\
\hline
\hfil Masked Familiar & \hfil Masked Unfamiliar & \hfil 0.08 & \hfil [-0.01, 0.15]\\
\hline
\end{tabular}
\end{adjustbox}
\caption{Result of post-hoc Tukey's test on Performance Accuracy}
\label{tab:ranova_PA}
\index{tables}
\end{table}

\begin{table}[H]
\centering
\begin{adjustbox}{width=\textwidth}
\begin{tabular}{|m{40mm}|m{40mm}|m{25mm}|m{25mm}|}
\hline
\hline
\multicolumn{4}{|c|}{\textbf{Dependent Variable: Reaction Time}}\\
\hline
\hline
\hfil \textbf{Group 1} & \hfil \textbf{Group 2} & \hfil \textbf{$p$-value} & \hfil \textbf{95\% C. I.}\\
\hline
\hfil Unmasked Famous & \hfil Masked Famous & \hfil \cellcolor{yellow!50}<0.001  & \hfil [-0.09, -0.03]\\
\hline
\hfil Unmasked Familiar & \hfil Masked Familiar & \hfil 0.33 & \hfil [-0.05, 0.02]\\
\hline
\hfil Unmasked Unfamiliar & \hfil Masked Unfamiliar & \hfil \cellcolor{yellow!50}0.01 & \hfil [-0.07, -0.01]\\
\hline
\hfil Unmasked Famous & \hfil Unmasked Familiar & \hfil 1 & \hfil [-0.03, 0.04]\\
\hline
\hfil Unmasked Famous & \hfil Unmasked Unfamiliar & \hfil \cellcolor{yellow!50}0.01 & \hfil [-0.09, -0.01]\\
\hline
\hfil Unmasked Familiar & \hfil Unmasked Unfamiliar & \hfil \cellcolor{yellow!50}0.003 & \hfil [-0.09, -0.02]\\
\hline
\hfil Masked Famous & \hfil Masked Familiar & \hfil \cellcolor{yellow!50}0.04 & \hfil [0.002, 0.08]\\
\hline
\hfil Masked Famous & \hfil Masked Unfamiliar & \hfil 0.16 & \hfil [-0.07, 0.01]\\
\hline
\hfil Masked Familiar & \hfil Masked Unfamiliar & \hfil \cellcolor{yellow!50}<0.001 & \hfil [-0.11, -0.04]\\
\hline
\end{tabular}
\end{adjustbox}
\caption{Result of post-hoc Tukey's test on Reaction Time}
\label{tab:ranova_RT}
\index{tables}
\end{table}

\begin{table}[hbt!]
\centering
\begin{adjustbox}{width=0.8\textwidth}
\begin{tabular}{|c|c|c|}
\hline
\hline
\multicolumn{3}{|c|}{\textbf{Dependent Variable: False-Positive Response Proportion}}\\
\hline
\hline
\textbf{Factor} & \textbf{F statistic} & \textbf{$p$-value}\\
\hline
Familiarity & 93.77 & \cellcolor{yellow!50}<0.0001\\
\hline
Mask & 33 & \cellcolor{yellow!50}<0.0001\\
\hline
Interaction & 0.41 & 0.66\\
\hline
\end{tabular}
\end{adjustbox}
\caption{Two-way repeated measures ANOVA on proportion of false-positive responses}\label{tab:anova_FP}
\index{tables}
\end{table}

\begin{table}[H]
\centering
\begin{adjustbox}{width=\textwidth}
\begin{tabular}{|m{40mm}|m{40mm}|m{25mm}|m{25mm}|}
\hline
\hline
\multicolumn{4}{|c|}{\textbf{Dependent Variable: False-Positive Response Proportion}}\\
\hline
\hline
\hfil \textbf{Group 1} & \hfil \textbf{Group 2} & \hfil \textbf{$p$-value} & \hfil \textbf{95\% C. I.}\\
\hline
\hfil Unmasked Famous & \hfil Masked Famous & \hfil \cellcolor{yellow!50}<0.0001  & \hfil [-0.02, -0.01]\\
\hline
\hfil Unmasked Familiar & \hfil Masked Familiar & \hfil \cellcolor{yellow!50}<0.0001 & \hfil [-0.03, -0.01]\\
\hline
\hfil Unmasked Unfamiliar & \hfil Masked Unfamiliar & \hfil \cellcolor{yellow!50}0.006 & \hfil [-0.03, -0.01]\\
\hline
\hfil Unmasked Famous & \hfil Unmasked Unfamiliar & \hfil \cellcolor{yellow!50}<0.0001 & \hfil [-0.06, -0.04]\\
\hline
\hfil Unmasked Familiar & \hfil Unmasked Unfamiliar & \hfil \cellcolor{yellow!50}<0.0001 & \hfil [-0.05, -0.03]\\
\hline
\hfil Masked Famous & \hfil Masked Unfamiliar & \hfil \cellcolor{yellow!50}<0.0001 & \hfil [-0.07, -0.04]\\
\hline
\hfil Masked Familiar & \hfil Masked Unfamiliar & \hfil \cellcolor{yellow!50}<0.0001 & \hfil [-0.06, -0.03]\\
\hline
\end{tabular}
\end{adjustbox}
\caption{Result of post-hoc Tukey's test on proportion of false-positive responses}
\label{tab:ranova_FP}
\index{tables}
\end{table}

\subsection*{Neural Results}

\subsubsection*{One way RM Anova}
\begin{table}[hbt!]
\centering
\begin{adjustbox}{width=0.7\textwidth}
\begin{tabular}{|c|c|c|}
\hline
\hline
\multicolumn{3}{|c|}{\textbf{Dependent Variable: N170 ERP amplitude}}\\
\hline
\hline
\textbf{Treatment Conditions} & \textbf{F statistic} & \textbf{$p$-value}\\
\hline
Unmasked: Famous, Familiar, Unfamiliar & 2.33 & 0.11\\
\hline
Masked: Famous, Familiar, Unfamiliar & 2.26 & 0.12\\
\hline
Famous: Unmasked, Masked & 1.28 & 0.27\\
\hline
Familiar: Unmasked, Masked & 4.86 & \cellcolor{yellow!50}0.04\\
\hline
Unfamiliar: Unmasked, Masked & 0.18 & 6.75\\
\hline
\end{tabular}
\end{adjustbox}
\caption{One-way repeated measures ANOVA on N170 ERP amplitude }\label{tab:one_anova_N170_all}
\index{tables}
\end{table}

\begin{table}[hbt!]
\centering
\begin{adjustbox}{width=0.7\textwidth}
\begin{tabular}{|c|c|c|}
\hline
\hline
\multicolumn{3}{|c|}{\textbf{Dependent Variable: N170 ERP amplitude, Familiar Face}}\\
\hline
\hline
\textbf{Factor} & \textbf{F statistic} & \textbf{$p$-value}\\
\hline
Mask & 4.75 & \cellcolor{yellow!50}0.044\\
\hline
Hemisphere & 0.07 & 0.8\\
\hline
Interaction & 0.0005 & 0.98\\
\hline
\end{tabular}
\end{adjustbox}
\caption{Two-way repeated measures ANOVA on N170 ERP amplitude for Familiar Face using hemisphere and mask as two factors}\label{tab:anova_N170_2_hemi_familiar}
\index{tables}
\end{table}

\begin{table}[hbt!]
\centering
\begin{adjustbox}{width=0.7\textwidth}
\begin{tabular}{|c|c|c|}
\hline
\hline
\multicolumn{3}{|c|}{\textbf{Dependent Variable: N250 ERP amplitude}}\\
\hline
\hline
\textbf{Treatment Conditions} & \textbf{F statistic} & \textbf{$p$-value}\\
\hline
Unmasked: Famous, Familiar, Unfamiliar & 3.26 & 0.051\\
\hline
Masked: Famous, Familiar, Unfamiliar & 2.78 & 0.08\\
\hline
Famous: Unmasked, Masked & 4.58 & \cellcolor{yellow!50}0.047\\
\hline
Familiar: Unmasked, Masked & 6.55 & \cellcolor{yellow!50}0.02\\
\hline
Unfamiliar: Unmasked, Masked & 1.3 & 0.27\\
\hline
\end{tabular}
\end{adjustbox}
\caption{One-way repeated measures ANOVA on N250 ERP amplitude }\label{tab:one_anova_N250_all}
\index{tables}
\end{table}


\begin{table}[hbt!]
\centering
\begin{adjustbox}{width=0.7\textwidth}
\begin{tabular}{|c|c|c|}
\hline
\hline
\multicolumn{3}{|c|}{\textbf{Dependent Variable: N250 ERP amplitude, Familiar Face}}\\
\hline
\hline
\textbf{Factor} & \textbf{F statistic} & \textbf{$p$-value}\\
\hline
Mask & 6.55 & \cellcolor{yellow!50}0.02\\
\hline
Hemisphere & 0.82 & 0.38\\
\hline
Interaction & 0.16 & 0.7\\
\hline
\end{tabular}
\end{adjustbox}
\caption{Two-way repeated measures ANOVA on N250 ERP amplitude for Familiar Face using Mask and Hemisphere as two factors}\label{tab:anova_N250_2_hemi_familiar}
\index{tables}
\end{table}

\begin{table}[H]
\centering
\begin{adjustbox}{width=\textwidth}
\begin{tabular}{|m{40mm}|m{40mm}|m{25mm}|m{25mm}|}
\hline
\hline
\multicolumn{4}{|c|}{\textbf{Dependent Variable: N250 ERP amplitude, Familiar Face}}\\
\hline
\hline
\hfil \textbf{Group 1} & \hfil \textbf{Group 2} & \hfil \textbf{$p$-value} & \hfil \textbf{95\% C. I.}\\
\hline
\hfil Right-hemisphere Unmasked & \hfil Right-hemisphere Masked & \hfil \cellcolor{yellow!50}0.01  & \hfil [-4.4, -0.6]\\
\hline
\end{tabular}
\end{adjustbox}
\caption{Result of post-hoc Tukey's test on N250 ERP amplitude for Familiar Face}
\label{tab:ranova_N250_2_hemi_familiar}
\index{tables}
\end{table}


\begin{table}[hbt!]
\centering
\begin{adjustbox}{width=0.7\textwidth}
\begin{tabular}{|c|c|c|}
\hline
\hline
\multicolumn{3}{|c|}{\textbf{Dependent Variable: N250 ERP amplitude, Famous Face}}\\
\hline
\hline
\textbf{Factor} & \textbf{F statistic} & \textbf{$p$-value}\\
\hline
Mask & 4.58 & \cellcolor{yellow!50}0.047\\
\hline
Hemisphere & 0.07 & 0.8\\
\hline
Interaction & 0.007 & 0.93\\
\hline
\end{tabular}
\end{adjustbox}
\caption{Two-way repeated measures ANOVA on N250 ERP amplitude for Famous Face using mask and hemishere as two factors }\label{tab:anova_N250_2_hemi_famous}
\index{tables}
\end{table}

\begin{table}[H]
\centering
\begin{adjustbox}{width=\textwidth}
\begin{tabular}{|m{40mm}|m{40mm}|m{25mm}|m{25mm}|}
\hline
\hline
\multicolumn{4}{|c|}{\textbf{Dependent Variable: N250 ERP amplitude, Famous Face}}\\
\hline
\hline
\hfil \textbf{Group 1} & \hfil \textbf{Group 2} & \hfil \textbf{$p$-value} & \hfil \textbf{95\% C. I.}\\
\hline
\hfil Right-hemisphere Unmasked & \hfil Right-hemisphere Masked & \hfil \cellcolor{yellow!50}0.048  & \hfil [0.01, 3.4]\\
\hline
\end{tabular}
\end{adjustbox}
\caption{Result of post-hoc Tukey's test on N250 ERP amplitude for Famous Face }
\label{tab:ranova_N250_2_hemi_famous}
\index{tables}
\end{table}

\subsubsection*{Two-way RM ANOVA results (Factors: Familiarity and Mask)}

Two way repeated measures ANOVA was done on electrode clusters corresponding to the following ERP components:
\\
\textbf{N170}

\begin{table}[hbt!]
\centering
\begin{adjustbox}{width=0.7\textwidth}
\begin{tabular}{|c|c|c|}
\hline
\hline
\multicolumn{3}{|c|}{\textbf{Dependent Variable: N170 ERP amplitude}}\\
\hline
\hline
\textbf{Factor} & \textbf{F statistic} & \textbf{$p$-value}\\
\hline
Familiarity & 0.18 & 0.84\\
\hline
Mask & 0.14 & 0.71\\
\hline
Interaction & 3.75 & \cellcolor{yellow!50}0.03\\
\hline
\end{tabular}
\end{adjustbox}
\caption{Two-way repeated measures ANOVA on N170 ERP amplitude (N=18)}\label{tab:anova_N170_2_all}
\index{tables}
\end{table}

Post-hoc Test:

\begin{table}[H]
\centering
\begin{adjustbox}{width=\textwidth}
\begin{tabular}{|m{40mm}|m{40mm}|m{25mm}|m{25mm}|}
\hline
\hline
\multicolumn{4}{|c|}{\textbf{Dependent Variable: N170 ERP amplitude}}\\
\hline
\hline
\hfil \textbf{Group 1} & \hfil \textbf{Group 2} & \hfil \textbf{$p$-value} & \hfil \textbf{95\% C. I.}\\
\hline
\hfil Unmasked Familiar & \hfil Masked Familiar & \hfil \cellcolor{yellow!50}0.041  & \hfil [-4.23, -0.1]\\
\hline
\end{tabular}
\end{adjustbox}
\caption{Result of post-hoc Tukey's test on N170 ERP amplitude }
\label{tab:ranova_N170_2_all}
\index{tables}
\end{table}

No other significant effects were found on post-hoc analysis (Tukey hsd)
\\
\textbf{N250}

\begin{table}[hbt!]
\centering
\begin{adjustbox}{width=0.7\textwidth}
\begin{tabular}{|c|c|c|}
\hline
\hline
\multicolumn{3}{|c|}{\textbf{Dependent Variable: N250 ERP amplitude}}\\
\hline
\hline
\textbf{Factor} & \textbf{F statistic} & \textbf{$p$-value}\\
\hline
Familiarity & 0.25 & 0.78\\
\hline
Mask & 0.84 & 0.37\\
\hline
Interaction & 7.34 & \cellcolor{yellow!50}0.002\\
\hline
\end{tabular}
\end{adjustbox}
\caption{Two-way repeated measures ANOVA on N250 ERP amplitude }\label{tab:anova_N250_2_all}
\index{tables}
\end{table}

Post-hoc Test:

\begin{table}[H]
\centering
\begin{adjustbox}{width=\textwidth}
\begin{tabular}{|m{40mm}|m{40mm}|m{25mm}|m{25mm}|}
\hline
\hline
\multicolumn{4}{|c|}{\textbf{Dependent Variable: N250 ERP amplitude}}\\
\hline
\hline
\hfil \textbf{Group 1} & \hfil \textbf{Group 2} & \hfil \textbf{$p$-value} & \hfil \textbf{95\% C. I.}\\
\hline
\hfil Unmasked Familiar & \hfil Masked Familiar & \hfil \cellcolor{yellow!50}0.02  & \hfil [-4.24, -0.41]\\
\hline
\hfil Unmasked Famous & \hfil Masked Famous & \hfil \cellcolor{yellow!50}0.047  & \hfil [0.02, 3.32]\\
\hline
\hfil Masked Familiar & \hfil Masked Famous & \hfil \cellcolor{yellow!50}0.01 & \hfil [0.48, 3.51]\\
\hline
\end{tabular}
\end{adjustbox}
\caption{Result of post-hoc Tukey's test on N250 ERP amplitude}
\label{tab:ranova_N250_2_all}
\index{tables}
\end{table}

No other significant effects were found on post-hoc analysis (Tukey hsd).

\textbf{P100}

\begin{table}[hbt!]
\centering
\begin{adjustbox}{width=0.7\textwidth}
\begin{tabular}{|c|c|c|}
\hline
\hline
\multicolumn{3}{|c|}{\textbf{Dependent Variable: P100 ERP amplitude}}\\
\hline
\hline
\textbf{Factor} & \textbf{F statistic} & \textbf{$p$-value}\\
\hline
Familiarity & 1.34 & 0.28\\
\hline
Mask & 0.08 & 0.78\\
\hline
Interaction & 2.57 & 0.09\\
\hline
\end{tabular}
\end{adjustbox}
\caption{Two-way repeated measures ANOVA on P100 ERP amplitude }\label{tab:anova_P100_2_all}
\index{tables}
\end{table}


\begin{table}[hbt!]
\centering
\begin{adjustbox}{width=0.7\textwidth}
\begin{tabular}{|c|c|c|}
\hline
\hline
\multicolumn{3}{|c|}{\textbf{Dependent Variable: P100 ERP amplitude (RH)}}\\
\hline
\hline
\textbf{Factor} & \textbf{F statistic} & \textbf{$p$-value}\\
\hline
Familiarity & 0.79 & 0.46\\
\hline
Mask & 0.48 & 0.5\\
\hline
Interaction & 1.77 & 0.19\\
\hline
\end{tabular}
\end{adjustbox}
\caption{Two-way repeated measures ANOVA on P100 ERP amplitude, right hemisphere cluster (N=17)}\label{tab:anova_P100_2_R}
\index{tables}
\end{table}


\begin{table}[hbt!]
\centering
\begin{adjustbox}{width=0.7\textwidth}
\begin{tabular}{|c|c|c|}
\hline
\hline
\multicolumn{3}{|c|}{\textbf{Dependent Variable: P100 ERP amplitude (LH)}}\\
\hline
\hline
\textbf{Factor} & \textbf{F statistic} & \textbf{$p$-value}\\
\hline
Familiarity & 1.33 & 028\\
\hline
Mask & 0.02 & 0.9\\
\hline
Interaction & 2.16 & 0.13\\
\hline
\end{tabular}
\end{adjustbox}
\caption{Two-way repeated measures ANOVA on P100 ERP amplitude, left hemisphere cluster}\label{tab:anova_P100_2_L}
\index{tables}
\end{table}

P100 component shows no significant main effect between familiarity and mask components and the same result holds when tested in separate hemispheres.

\printbibliography
\end{document}